# Tactile Internet and its Contribution in the Development of Smart Cities

Tanweer Alam[a]

[a]Department of Computer Science, Islamic University of Madinah, Saudi Arabia Email: tanweer03@iu.edu.sa

**Abstract**

The Tactile Internet (TI) is an emerging technology next to the Internet of Things (IoT). It is a revolution to develop the smart cities, communities and cultures in the future. This technology will allow the real-time interaction between human and machines as well as machine-to-machine with the 1ms challenge to achieve in round trip latency. The term TI is defined by International Telecommunication Union (ITU) in August 2014. The TI provides fast, reliable, secure and available internet network that is the requirements of the smart cities in 5G. Tactile internet can develop the part of world where the machines are strong and human are weak. It increases the power of machines so that the value of human power will increase automatically. In this framework, we have presented the idea of tactile internet for the next generation smart cities. This research will provide a high-performance reliable framework for the internet of smart devices to communicate with each other in a real-time (1ms round trip) using IEEE 1918.1 standard.  The objective of this research is expected to bring a new dimension in the research of the smart cities.

## Introduction:

The proposed research entitled "Tactile Internet based reliable communication framework for Smart cities in 5G" is a step forward in wireless networking and IoT where we propose new reliable framework based on Tactile Internet. The Wireless communication is the key of Internet of things and Tactile Internet. It is expected to exceed 50 billion connected devices by 2020 and most of these nodes cannot be connected by wireline. In order to enable critical applications such as smart factories or smart buildings, the networking protocols have to deal with the non-deterministic nature of wireless links. In the 5th generation communication system, the secure and reliable data packets will rely on the network with high availability and low latency.

5G enables the dynamic control of nodes and low latency. Ultra-reliability feature is an interface working with high availability and low latency in Tactile Internet that brings in fifth generation networks. The tactile internet will act in the area of societies so that it required ultra-reliability feature to empower the peoples as well as machines for collaborating with their neighbors. The connections among ultra-reliability networks are extremely hard situation to keep low packets corruption. The tactile internet will provide a platform for measuring, controlling, monitoring and scaling the smart objects in reality or virtually in the smart cities.  The ultra-low latency, reliability and availability in controlling are the main features of tactile internet that makes it advanced in 5G.

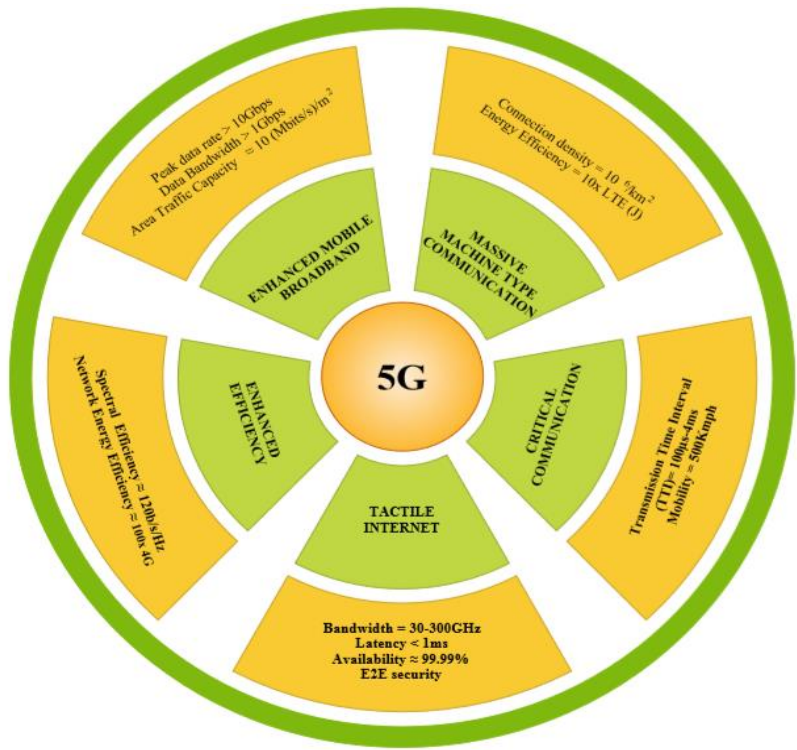

Fig. 1: 5G Network

Because of these features the internet want to move from mobile to Tactile Internet. The proposed research work in this project is an enhancement and implementation of reliable framework based on Tactile Internet emerging technology next to the internet of things. The research outcome is to establish a new reliable framework for communication among human to machine and machine to machine in the future smart cities. The proposed research uses the correct and efficient simulation of a desired study and can be implemented in a framework of smart cities. In the future, researchers can enhance this research and implement in the internet of everything framework.

The objective of this research is to create the new reliable communication framework for the smart cities using the Tactile Internet a next revolution of internet of things. This research is based on low-latency, ultra-high availability and high-performance concepts of Tactile Internet. The framework provides QoS through reducing the latency (1ms in round trip) also the variety of the quantity of smart devices. In this research we consider idle state in order to makes our examination more efficient, at that point the general execution regarding the overall performance of the framework is evaluated. The framework will monitor and analyze the real-time data collected from network and then taking the action.

The following are the key points-

The research is primarily focused on next generation Internet for smart cities. It enables smart devices to communicate with another device among internet of smart devices using fast, reliable and secure tactile internet. The proposed framework for communication will access across the internet of smart devices. The results of proposed research will be compared with previous study in the same area.

## 1. Literature Review

In 1991, Theodore S. Rappaport published an article entitles "The wireless revolution", in this paper he presented the wireless communications is the emerging technology as a key for communication among human as well as devices [1]. In the med of 2006, Amazon achieved a prominent milestone by testing elastic computing cloud (EC 2) which initialized the spark of cloud computing in it. However, the term cloud computing has not coned until March 2007 [2]. The following year brought even more rapid development of the newly emerged paradigm. Furthermore, the cloud computing infrastructure services have widened to include (SaaS) software as a service. In the mid of 2012, oracle cloud has been introduced, where it supports different deployment models. It is provisioned as the first unified collection of its solutions which is under continues developments. Nowadays, typing a cloud computing in any search engine will result in a tremendous result. For example, it would result more than 139,000,000 matches in Google.

Table.2: Comparison of peak data rate and latency [3]

| Technology | Peak Data Rate | Latency |
|---|---|---|
| GPRS | 114 kbps | ~ 500ms |
| EDGE | 236.8 kbps | ~ 250ms |
| W-CDMA | 384 kbps | ~ 200ms |
| HSPA | 2 Mbps | ~ 150ms |
| HSPA | 42 Mbps | ~ 70ms |
| LTE | 300 Mbps | ~ 30ms |
| LTE-A | 1 Gbps | ~ 20ms |

In 2019, Ishan Budhiraja, et. al. published a paper [4], In this paper the authors were presented the tactile internet for smart communities in 5G. They summarize the use of non-orthogonal multiple access protocol in 5G. In the technical report [5], the authors represented the tactile internet as the next revolution after the Internet of things. In article [6], the cloud-based queuing model is explained for the tactile internet. In article [7], the author discussed about haptic communication system. In article [8], the authors are enabled the tactile internet for ultra-reliability and fast response time (<1ms). In the paper [9], the authors present a review on tactile internet for industries. It represents the role of tactile internet in the future industries. In the thesis [10], the author explored the challenges and standards for the tactile internet in 5G. In [11], the 5G-based Tactile internet framework is designed. Very few articles are written on Tactile internet. The previous studies showed us the role, use of tactile internet in 5th generation.

## 2. Research Importance

The smart devices are increasing exponentially day by day in the whole world. They provide much more facility to the end users and also attach with their daily life. Smart devices can connect to the internet easily for sending and receiving data within the network. The smart devices are not just smart phones, it may be smart refrigerator, Smart home automation entry point, smart air conditioners, Smart hubs, Smart thermostat, Color changing smart LEDs, Smart Watches and smart Tablets etc. in internet of things framework they are connected to each other through internet.

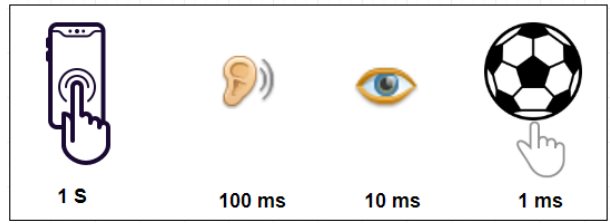

Fig. 2: Human reaction time

Table 1. IoT Devices installed category and year wise (in Millions)

| Category | 2016 | 2017 | 2018 | 2020 |
|---|---|---|---|---|
| IoT Devices | 3963.0 | 5244.3 | 7036.3 | 12863.0 |
| Business: Across Industries | 1102.1 | 1501.0 | 2132.6 | 4381.4 |
| Business: Vertical specific | 1316.6 | 1635.4 | 2027.7 | 3171.0 |
| Total | 6381.8 | 8380.6 | 11196.6 | 20415.4 |

The proposed research plan builds research on extending the performance of communication in internet of things using tactile internet. The transfer data from one configuration to another using wireless networks starts from 1973 in the form of packets radio network. They were able to communicate with another same configuration devices. Recent work is continuing on a project called the Serval Project. It provides networks facility to android devices for communication in infrastructure less network. Whereas our research is concerned about the high-performance communication in internet of smart devices for smart cities. The main contribution of this research is the creation of the reliable communication framework and provide secure, reliable and fast communication using Tactile Internet among the internet of smart devices. The previous studies have been focused on the creation and optimization the framework for communication, but such research doesn't perform the full framework for secure and reliable communication among internet of smart devices for smart cities.

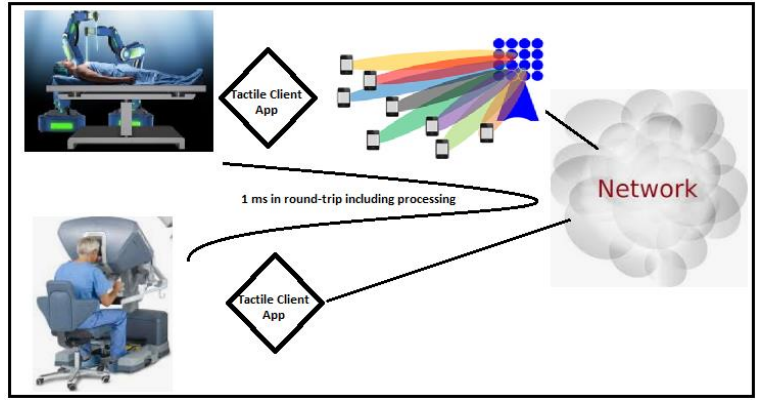

Fig. 3: An example of Tactile Internet

## 3. Conclusion

The main contribution of this research is designing a framework for ultra-reliable, low latency and high availability communication in Internet of smart devices for future smart cities using the Tactile Internet. The proposed framework is specifically appropriate for applications in which data is periodically transmitted in internet of smart devices environment. In these applications, on one hand, packets are being produced based on a certain periodic time pattern. On the other hand, service time is always a random variable with general distribution. Therefore, service time might temporarily exceed the period time which, as an inevitable consequence some packets might encounter a busy channel and be dropped. We solve this problem by proposing the new communication framework. We demonstrate that proposed reliable framework, not only increases the throughput, but also the direct connection between the generation (sensors) and communication packet systems are eliminated which make the system far more stable. Moreover, in order to enhance the proposed model, we have employed retransmission scheme, variable packet length, and saturated traffic condition. The solution of this research is summarized as follows. The implementation of proposed framework for communication among internet of smart devices in 5G will be programmed to execute on to the internet of things using Tactile Internet concepts. The idea will focus into three main concepts, these concepts are Reliability, Security and availability. The proposed study supports the wireless networking technology to establish a reliable framework among internet of devices for smart cities.

## References

[1]. Li, Shancang, Li Da Xu, and Shanshan Zhao. "The internet of things: a survey." Information Systems Frontiers 17.2 (2015): 243-259.

[2]. Rappaport, Theodore S. "The wireless revolution." IEEE Communications Magazine 29, no. 11 (1991): 52-71.
[3]. Garfinkel, Simson. "An evaluation of amazon's grid computing services: EC2, S3, and SQS." (2007).
[4]. M. Chowdhury and M. Maier, "Local and Non-Local Human-to-Robot Task Allocation in Fiber-Wireless Multi-Robot Networks," IEEE Systems Journal, vol. 12, no. 3, pp. 2250-2260, Sept. 2018
[5]. Budhiraja I, Tyagi S, Tanwar S, Kumar N, Rodrigues JJ. Tactile Internet for Smart Communities in 5G: An Insight for NOMA-based Solutions. IEEE Transactions on Industrial Informatics. 2019 Jan 14.
[6]. Aijaz, Adnan, Mischa Dohler, A. Hamid Aghvami, Vasilis Friderikos, and Magnus Frodigh. "Realizing the tactile internet: Haptic communications over next generation 5G cellular networks." IEEE Wireless Communications 24, no. 2 (2017): 82-89.
[7]. Fettweis, Gerhard, and Siavash Alamouti. "5G: Personal mobile internet beyond what cellular did to telephony." IEEE Communications Magazine 52, no. 2 (2014): 140-145.
[8]. Fettweis, Gerhard P. "The tactile internet: Applications and challenges." IEEE Vehicular Technology Magazine 9, no. 1 (2014): 64-70.
[9]. Ji, Hyoungju, Sunho Park, Jeongho Yeo, Younsun Kim, Juho Lee, and Byonghyo Shim. "Ultra-reliable and low-latency communications in 5G downlink: Physical layer aspects." IEEE Wireless Communications 25, no. 3 (2018): 124-130.
[10]. Aijaz A, Sooriyabandara M. The Tactile Internet for Industries: A Review [35pt]. Proceedings of the IEEE. 2018 Nov 21.
[11]. Cakuli, Julian. "Application Scenarios, Research Challenges and Standardization for Tactile Internet." (2017).
[12]. Li, Chong, Chih-Ping Li, Kianoush Hosseini, Soo Bum Lee, Jing Jiang, Wanshi Chen, Gavin Horn, Tingfang Ji, John E. Smee, and Junyi Li. "5G-based systems design for tactile Internet." Proceedings of the IEEE 99 (2018): 1-18.
[13]. Maier, Martin, Mahfuzulhoq Chowdhury, Bhaskar Prasad Rimal, and Dung Pham Van. "The tactile internet: vision, recent progress, and open challenges." IEEE Communications Magazine 54, no. 5 (2016): 138-145.
[14]. Oteafy, Sharief MA, and Hossam S. Hassanein. "Leveraging Tactile Internet Cognizance and Operation via IoT and Edge Technologies." Proceedings of the IEEE 107, no. 2 (2019): 364-375.
[15]. Simsek, Meryem, Adnan Aijaz, Mischa Dohler, Joachim Sachs, and Gerhard Fettweis. "5G-enabled tactile internet." IEEE Journal on Selected Areas in Communications 34, no. 3 (2016): 460-473.
[16]. Lethaby, Nick. "Wireless connectivity for the Internet of Things: One size does not fit all." Texas Instruments (2017).
[17]. Cao H. What is the next innovation after the internet of things?. arXiv preprint arXiv:1708.07160. 2017 Aug 23.
[18]. Gholipoor, Narges, Saeedeh Parsaeefard, Mohammad Reza Javan, Nader Mokari, and Hamid Saeedi. "Cloud-based Queuing Model for Tactile Internet in Next Generation of RAN." arXiv preprint arXiv:1901.09389 (2019).

[19]. Aijaz, Adnan. "Towards 5G-enabled tactile internet: Radio resource allocation for haptic communications." In Wireless Communications and Networking Conference Workshops (WCNCW), 2016 IEEE, pp. 145-150. IEEE, 2016. Alam T, Benaida M. "The Role of Cloud-MANET Framework in the Internet of Things (IoT)", International Journal of Online Engineering (iJOE). Vol. 14(12), pp. 97-111. DOI: https://doi.org/10.3991/ijoe.v14i12.8338

[20]. Alam, Tanweer. "Middleware Implementation in Cloud-MANET Mobility Model for Internet of Smart Devices", International Journal of Computer Science and Network Security, 17(5), 2017. Pp. 86-94

[21]. Alam T, Benaida M. CICS: Cloud–Internet Communication Security Framework for the Internet of Smart Devices. International Journal of Interactive Mobile Technologies (iJIM). 2018 Nov 1;12(6):74-84. DOI: https://doi.org/10.3991/ijim.v12i6.6776

[22]. Tanweer Alam, "Blockchain and its Role in the Internet of Things (IoT)", International Journal of Scientific Research in Computer Science, Engineering and Information Technology, vol. 5(1), pp. 151-157, 2019. DOI: https://doi.org/10.32628/CSEIT195137

[23]. Tanweer Alam, Baha Rababah, "Convergence of MANET in Communication among Smart Devices in IoT", International Journal of Wireless and Microwave Technologies(IJWMT), Vol.9, No.2, pp. 1-10, 2019. DOI: 10.5815/ijwmt.2019.02.01

[24]. Tanweer Alam, "IoT-Fog: A Communication Framework using Blockchain in the Internet of Things", International Journal of Recent Technology and Engineering (IJRTE), Volume-7, Issue-6, 2019.

[25]. Alam, Tanweer, and Mohammed Aljohani. "Design and implementation of an Ad Hoc Network among Android smart devices." In Green Computing and Internet of Things (ICGCIoT), 2015 International Conference on, pp. 1322-1327. IEEE, 2015. DOI: https://doi.org/10.1109/ICGCIoT.2015.7380671

[26]. Alam, Tanweer, and Mohammed Aljohani. "An approach to secure communication in mobile ad-hoc networks of Android devices." In 2015 International Conference on Intelligent Informatics and Biomedical Sciences (ICIIBMS), pp. 371-375. IEEE, 2015. DOI: https://doi.org/10.1109/iciibms.2015.7439466

[27]. Aljohani, Mohammed, and Tanweer Alam. "An algorithm for accessing traffic database using wireless technologies." In Computational Intelligence and Computing Research (ICCIC), 2015 IEEE International Conference on, pp. 1-4. IEEE, 2015. DOI: https://doi.org/10.1109/iccic.2015.7435818

[28]. Alam, Tanweer, and Mohammed Aljohani. "Design a new middleware for communication in ad hoc network of android smart devices." In Proceedings of the Second International Conference on Information and Communication Technology for Competitive Strategies, p. 38. ACM, 2016. DOI: https://doi.org/10.1145/2905055.2905244

[29]. Alam, Tanweer. (2018) "A reliable framework for communication in internet of smart devices using IEEE 802.15.4." ARPN Journal of Engineering and Applied Sciences 13(10), 3378-3387.

[30]. Tanweer Alam, "A Reliable Communication Framework and Its Use in Internet of Things (IoT)", International Journal of Scientific Research in Computer Science, Engineering and Information

Technology (IJSRCSEIT), Volume 3, Issue 5, pp.450-456, May-June.2018 URL: http://ijsrcseit.com/CSEIT1835111.

[31]. Alam, Tanweer. "Fuzzy control based mobility framework for evaluating mobility models in MANET of smart devices." ARPN Journal of Engineering and Applied Sciences 12, no. 15 (2017): 4526-4538.

[32]. Alam, Tanweer, Arun Pratap Srivastava, Sandeep Gupta, and Raj Gaurang Tiwari. "Scanning the Node Using Modified Column Mobility Model." Computer Vision and Information Technology: Advances and Applications 455 (2010).

[33]. Alam, Tanweer, Parveen Kumar, and Prabhakar Singh. "SEARCHING MOBILE NODES USING MODIFIED COLUMN MOBILITY MODEL.", International Journal of Computer Science and Mobile Computing, (2014).

[34]. Alam, Tanweer, and B. K. Sharma. "A New Optimistic Mobility Model for Mobile Ad Hoc Networks." International Journal of Computer Applications 8.3 (2010): 1-4. DOI: https://doi.org/10.5120/1196-1687

[35]. Singh, Parbhakar, Parveen Kumar, and Tanweer Alam. "Generating Different Mobility Scenarios in Ad Hoc Networks.", International Journal of Electronics Communication and Computer Technology, 4(2), 2014

[36]. Sharma, Abhilash, Tanweer Alam, and Dimpi Srivastava. "Ad Hoc Network Architecture Based on Mobile Ipv6 Development." Advances in Computer Vision and Information Technology (2008): 224.

[37]. Aljohani, Mohammed, and Tanweer Alam. "Real Time Face Detection in Ad Hoc Network of Android Smart Devices." Advances in Computational Intelligence: Proceed-ings of International Conference on Computational Intelligence 2015. Springer Singa-pore, 2017.DOI: https://doi.org/10.1007/978-981-10-2525-9_24

[38]. M. Aljohani and T. Alam, "Design an M-learning framework for smart learning in ad hoc network of Android devices," 2015 IEEE International Conference on Computational Intelligence and Computing Research (ICCIC), Madurai, 2015, pp. 1- 5. https://doi.org/10.1109/ICCIC.2015.7435817